\documentclass[journal]{IEEEtran}
\usepackage{algorithm,algpseudocodex,amsmath,amssymb}
\usepackage{cite,graphicx,subfig,xcolor,array,balance}
\usepackage{tikz,nicematrix}
\usetikzlibrary{fit}
\newcommand{\Upd}[1]{#1}
\newcommand{\Edt}[1]{#1}
\newcommand{\Rev}[1]{{#1}}

\title{Enhancing Accuracy-Privacy Trade-off in Differentially Private Split Learning}
\author{Ngoc Duy Pham, Khoa T. Phan,~\IEEEmembership{Member,~IEEE,} and Naveen Chilamkurti,~\IEEEmembership{Senior Member,~IEEE,}
\thanks{© 2024 IEEE.  Personal use of this material is permitted.  Permission from IEEE must be obtained for all other uses, in any current or future media, including reprinting/republishing this material for advertising or promotional purposes, creating new collective works, for resale or redistribution to servers or lists, or reuse of any copyrighted component of this work in other works.}
\thanks{N. D. Pham was with the School of Computing, Engineering, and Mathematical Sciences, La Trobe University, Melbourne, Australia (e-mail: ngocduy.pham@latrobe.edu.au). He is currently with the Faculty of Information Technology, School of Technology, Van Lang University, Ho Chi Minh City, Vietnam (e-mail: duy.pn@vlu.edu.vn).}
\thanks{K. T. Phan and N. Chilamkurti are with the School of Computing, Engineering, and Mathematical Sciences (SCEMS), La Trobe University, Melbourne, VIC 3086, Australia (e-mail: k.phan@latrobe.edu.au, n.chilamkurti@latrobe.edu.au). {\it{Corresponding authors: K. T. Phan and N. Chilamkurti}}.}}
\begin{document}
\maketitle
\begin{abstract}
Split learning (SL) aims to protect user data privacy by distributing deep models between the client-server and keeping private data locally. Only processed or `smashed' data can be transmitted from the clients to the server during the SL process. However, recently proposed model inversion attacks can recover original data from smashed data. To enhance privacy protection against such attacks, one strategy is to adopt differential privacy (DP), which involves safeguarding the smashed data at the expense of some accuracy loss. This paper presents the first investigation into the impact on accuracy when training multiple clients in SL with various privacy requirements. Subsequently, we propose an approach that reviews the DP noise distributions of other clients during client training to address the identified accuracy degradation. We also examine the application of DP to the local model of SL to gain insights into the trade-off between accuracy and privacy. Specifically, the findings reveal that introducing noise in the later local layers offers the most favorable balance between accuracy and privacy. Drawing from our insights in the shallower layers, we propose an approach to reduce the size of smashed data to minimize data leakage while maintaining higher accuracy, optimizing the accuracy-privacy trade-off. Additionally, smashed data of a smaller size reduces communication overhead on the client side, mitigating one of the notable drawbacks of SL. Intensive experiments on various datasets demonstrate that our proposed approaches provide an optimal trade-off for incorporating DP into SL, ultimately enhancing the training accuracy for multi-client SL with varying privacy requirements.
\end{abstract}


\section{Introduction}
Privacy concerns are prevalent across diverse domains, spanning healthcare, finance, e-commerce, and more. The concerns often impose limitations on sharing the raw data essential for training accurate deep neural networks (DNNs), also referred to as deep models \cite{PrivacySurvey20Fatemehsadat,SurveyPPFL21Yin}. Highly personal user data, such as medical or financial records, fall within the category of sensitive information, making the reluctance to share such data understandable. This situation underscores the importance of privacy-preserving deep learning (DL) frameworks, which aim to leverage the potential of DL applications while respecting privacy constraints. For instance, sensitive data can be employed to develop DL applications providing insights into areas such as predicting rare diseases or identifying financial irregularities. However, the challenge lies in achieving these objectives without direct access to the raw data. Various techniques can be employed, including anonymization, obfuscation, or data encryption. However, a persistent concern is the possibility of decrypting the original data from the trained model through adversarial attacks \cite{SurveyPPFL21Yin}. Therefore, it is crucial to facilitate the collaborative training of distributed machine learning models without requiring data sharing.

A recent approach known as split learning (SL) \cite{SplitLearning18Gupta,SL4Health18Vepakomma} has emerged as a solution to address this challenge, enabling multiple parties to collaboratively train a model without explicitly sharing raw input data. In a standard two-party SL setup, the deep model is partitioned between two entities: the data owner, referred to as the client, and the computing service, known as the public server in the cloud. Throughout the training process, during the forward propagation phase, the client feeds its raw data into its designated model part. It then transmits the processed data, often termed `smashed data' (which is the output of the final layer, also called as the split layer, within the client part), to the server. Subsequently, the server continues the forward pass by further processing the received data through its respective model part. During the back-propagation phase, the gradients of the split layer are sent back to the client \cite{SplitLearning18Gupta}. This mechanism allows the client to train the collaborative model without exposing its private training data to the server.
\Upd{Moreover, this splitting manner, wherein only the initial layers are distributed to the client while the rest are deployed at the server, facilitates learning on low-end devices such as IoT, as the majority of the computational burden is offloaded to the server. This is in contrast to federated learning \cite{FederatedLearning17Jakub}, another distributed learning framework where the entire model is executed on the devices.}

\begin{figure}[t]
\centering
\includegraphics[width=.38\textwidth]{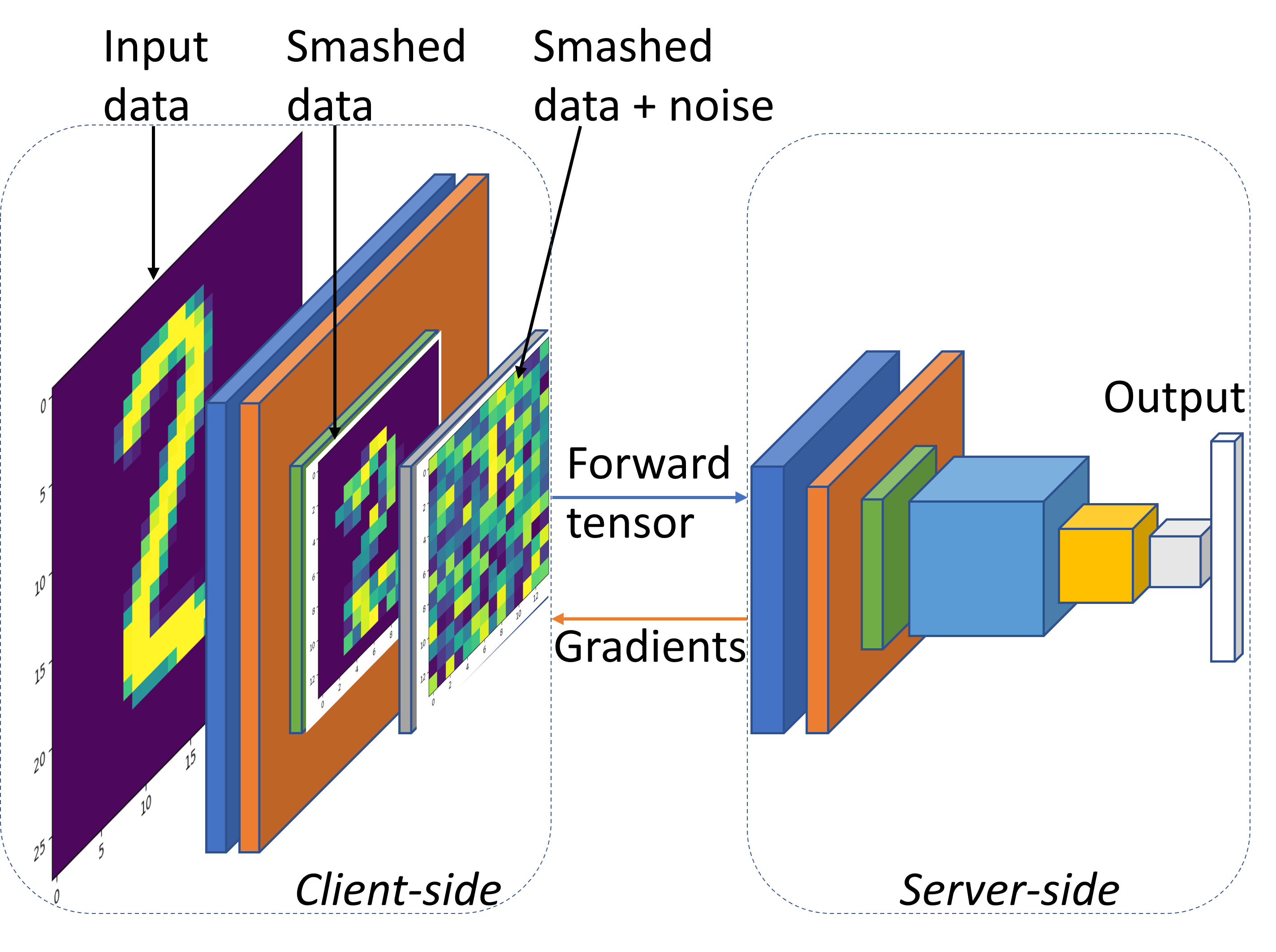}
\caption{SL with DP noise for privacy-preserving enhancement.}
\label{fig:SL_DP_noise}
\end{figure}

Unfortunately, merely sharing smashed data does not provide an absolute guarantee of protecting raw private data. The shared smashed data inherently contains substantial latent information about the input data and could potentially serve as a basis for relatively straightforward reversal techniques, such as visualization \cite{SLon1DCNN20Sharif,BinarizingSL23Pham}, or more intricate reconstructions like model inversion attacks \cite{AttackingProtecting22He,ModelInversion19He,UnSplit22Erdogan}. Several efforts \cite{SLon1DCNN20Sharif,BinarizingSL23Pham,AttackingProtecting22He,NotJustPrivacy18Wang,Shredder20Mireshghallah,PracticalDefences21Titcombe,CollaborativeInference22Ryu,CutMix4SL22Oh,CutMix4SL24Oh} aim to mitigate this risk through the implementation of differential privacy (DP), which is currently a standard practice in privacy-preserving machine learning. For example, in \cite{CollaborativeInference22Ryu}, the researchers explore the utilization of DP to safeguard users' data privacy during collaborative inference using SL for IoT. Fig. \ref{fig:SL_DP_noise} visually illustrates the integration of DP noise into smashed data to enhance data privacy preservation in the context of SL. A notable privacy vulnerability of SL arises from the fact that a neural network operates as a differentiable smooth function, inherently susceptible to functional inversion. Partitioning the model between clients and the server does not alter this characteristic \cite{CombinedFLSL22Duan}. Researchers propose countering this vulnerability by introducing DP noise to the smashed data, making individual identification no longer feasible. They assess the effectiveness of their approach across various image datasets, demonstrating that DP can effectively safeguard the privacy of users data while achieving a degree of accuracy reduction.

The application of DP to smashed data to safeguard user data can be extended to a multi-data-owner environment where multiple clients collaboratively conduct model training with a cloud server. In most cases, a uniform privacy level is considered for all clients. The recent work in \cite{InputDiscriminativeDP20Gu} explores the context of local DP, where each data item is associated with distinct privacy levels, and noisy data is sent to a public server to process frequency estimation tasks. The authors demonstrate that their approach outperforms other related local DP mechanisms, which require a unique privacy budget for all categorical data. However, both \cite{InputDiscriminativeDP20Gu} and its successor \cite{TuningPrivacyBudgets23}, which focus on tuning privacy budgets to minimize the risk of re-identification attacks, only evaluate the efficiency of their schemes based on the aggregated results of queries and do not consider the impact of varying privacy levels on the utility of the queries. Inspired by the concept of multi-level privacy, we investigate a scenario of multi-client SL in which some clients employ DP for additional privacy and then identify the accuracy degradation depending on the difference in privacy per client. To the best of our knowledge, this work is the first to investigate SL in multiple clients with different privacy levels, revealing that accuracy is susceptible to the inherent forgetting phenomenon of SL.
\Upd{The forgetting phenomenon \cite{Forgetting99French,EvaluationDML22Gao}, also known as catastrophic forgetting, occurs when the network (server model) is sequentially trained on multiple tasks (clients), leading to changes in weights crucial for one task to meet the objectives of another task.}
To address this concern, we propose a training approach that involves the server-side review of DP noise distributions to enhance accuracy.

In much of the existing literature, it is noteworthy that calibrated noise is primarily added exclusively to the outputs from the split layer of SL, with little consideration to applying DP to other local layers. In \cite{AttackingProtecting22He}, alongside their comprehensive privacy attack proposals, the authors explore the application of noise at the original input to defense, claiming its superior effectiveness compared to noise at the split layer. However, this claim lacks a deeper investigation or thorough evaluation. Nevertheless, a central challenge with DP lies in striking the right balance between preserving model utility and ensuring data privacy. Increased noise levels can provide stronger privacy guarantees, but inevitably impact the overall model utility (accuracy). Concerning SL with DP, a fundamental question arises: Within the SL framework, where should DP be applied to achieve the optimal trade-off between model accuracy and data privacy? To address this research question, we examine the trade-offs between accuracy and privacy when applying DP at different local layers of SL to identify the most effective position for introducing noise and achieving an improved trade-off. \Rev{Our findings suggest that applying DP noise further from the input layer achieves a better balance between privacy and accuracy. Conversely, at the same DP level, adding noise directly to the input layer more effectively reduces data leakage but significantly compromises model accuracy, resulting in a less favorable trade-off.}
Based on these insights, we propose a method to reduce the exposed (smashed) data size, limiting data leakage and reducing SL communication costs while maintaining the most favorable accuracy and privacy trade-offs. In summary, our contributions in this paper are as follows:

\begin{enumerate}
\item Identifying the influence of different privacy levels on the accuracy of multi-client SL and introducing a server-side approach that reviews noise distributions to address the forgetting phenomenon, thereby enhancing learning performance.
\item Analyzing the trade-offs between model accuracy and data privacy by applying DP at different local layers in SL, with our findings highlighting that introducing DP noise to the split layer yields the most favorable trade-off. We also propose a method to enhance the performance of SL with DP.
\item Conducting an extensive empirical evaluation with real-world datasets. The experiments validate the efficacy of our proposed approaches, effectively mitigating the forgetting effect and substantially improving learning performance, aligning it closely with the benchmark.
\end{enumerate}

The rest of the paper is structured as follows: Section \ref{background} provides background information on multi-client SL, model inversion attacks, and DP-based defense. In Section \ref{variousDP}, we present the proposed approach for addressing the identified accuracy degradation of multi-client SL considering different privacy requirements. The analysis of the trade-offs between model utility and data privacy when applying DP to SL is detailed in Section \ref{trade-off}, followed by the proposed approach to enhance the performance of SL with DP. Section \ref{experiment} reports the experiment results pertaining to the trade-off analysis and the efficacy of the proposed approaches. Finally, the paper concludes with insights and future research directions in Section \ref{conclusion}.

\section{Preliminaries}\label{background}
This section provides background knowledge about multi-client SL, with the current model inversion attacks and commonly employed defenses involving DP.

\subsection{Multi-client split learning}
The fundamental concept behind SL is to horizontally divide a learning model, denoted as $h_{\theta}=g_w\circ f_u$ where $\theta$, or $u$ and $w$ represent the model's parameters that need to be optimized during the model training process. The initial layers, referred to as $f_u$, are executed on client devices, while the remaining layers, designated as $g_w$, reside on the central server. 
During the training process, \Upd{at a given time $t$,} the $i$-th client ($C_i$) inputs its data, $x^t$, into its local model, carrying out forward propagation until reaching the split layer. The output, represented as the smashed data $z^t\triangleq f_{u_i}(x^t)$, along with the corresponding label, $y^t$, are then transmitted to the server. Subsequently, the server continues the forward propagation process and computes the derivative of the loss function, and then initiates back propagation until the split layer. The gradients associated with the split layer are returned to the client to complete the remaining back-propagation steps. Both the server and the client update their respective model parameters, $u_i$ and $w$, using the gradient descent algorithm. At the end of each iteration, client $C_i$ shares its updated model parameters, $u_i$, with the subsequent client, $C_{i+1}$, to repeat the aforementioned process. This cycle persists until all clients complete one full round of training. Such a complete global iteration is repeated until the model converges \cite{SplitLearning18Gupta,SL4Health18Vepakomma}.

\subsection{Model inversion attacks on SL}
The privacy preservation of SL, where user data remain local, faces potential vulnerabilities from the recently proposed model inversion attacks \cite{ModelInversion19He,AttackingProtecting22He}, which aim to reconstruct user raw data from the exposed smashed data. In a black-box setting, the authors of \cite{ModelInversion19He} introduce an inverse network as an effective means of accurately recovering raw data. The procedure involves the adversary generating a set of samples $X=\{x_1,x_2,\cdots,x_m\}$ of the same type as the training data. These samples are used to query the target system, resulting in corresponding smashed data denoted as $Z=\{f_u(x_i)|x_i\in X\}$. Subsequently, the inverse network can be directly trained using $Z$ as the training input and $X$ as the training output, guided by Eq. (\ref{eq:inverseNetwork}):
\begin{equation}
f^{-1}=arg\min_v\frac{1}{m}\sum^m_{i=1}||g_v\bigl(f_u(x_i)\bigr)-x_i||^2
\label{eq:inverseNetwork}
\end{equation}
\noindent Here, $g_v$ represents the inverse network subject to optimization \cite{ModelInversion19He}. Once the inverse network is established, the adversary can recover any raw data from the smashed data, effectively achieving $x\approx f^{-1}(z)$. In terms of defense, obfuscating smashed data with DP noise is one of the general approaches suggested by the authors.

\subsection{SL with differential privacy}
Differential privacy (DP) \cite{DifferentialPrivacy06Dwork} is a mathematical framework designed for privacy-preserving data analysis, defined as follows:

\textbf{Definition.} Given any two neighboring inputs $D$ and $D'$, which differ in only one data item, a mechanism $\mathcal{M}$ provides $(\epsilon,\delta)$-differential privacy if:
\begin{equation}
\Pr[\mathcal{M}(D)\in S]\leq e^{\epsilon}\cdot\Pr[\mathcal{M}(D')\in S]+\delta,
\end{equation}

\noindent where $S$ is any subset of $\mathcal{M}$'s image. Intuitively, it should be challenging to determine whether a given output is produced from $D$ or $D'$. The parameter $\epsilon$ is often referred to as the privacy budget, with a lower value indicating stronger privacy protection, \Upd{and the parameter $\delta$ is a bound on the external risk that is not constrained by $\epsilon$.} To establish DP, the typical approach involves introducing calibrated noise to the output of a function $f(\cdot)$, relying on specific probability distributions \cite{DifferentialPrivacy06Dwork}. In our study, we employ the widely used Gaussian distribution, denoted as $\mathcal{N}(\sigma^2)$, which represents Gaussian (normal) distribution centered at 0 with a variance of $\sigma^2$, as formulated below:
\begin{equation}
\mathcal{F}(x)=f(x)+\mathcal{N}(\sigma^2),
\label{eq:Gaussian}
\end{equation}
\[\text{where }\sigma^2=\frac{2s^2\log(1.25/\delta)}{\epsilon^2}.\]

$\mathcal{F}$ is said to provide $(\epsilon,\delta)$-DP, with $s$ representing the sensitivity of $f$. Our choice of the Gaussian mechanism is motivated by its ability to accommodate the summation of two distributions, which will be utilized in our subsequent proposal. Considering $f$ as the local model, the literature encompasses a substantial body of work \Rev{\cite{SLon1DCNN20Sharif,BinarizingSL23Pham,AttackingProtecting22He,NotJustPrivacy18Wang,Shredder20Mireshghallah,PracticalDefences21Titcombe,CollaborativeInference22Ryu,CutMix4SL22Oh,CutMix4SL24Oh}} dedicated to introducing DP noise to smashed data to mitigate the risk associated with raw data reconstruction.
\Rev{Additionally, due to the post-processing property of DP \cite{ProgrammingDP21Near}, noise can be applied dynamically to the split layer, input, or any local layer \emph{without} compromising the privacy guarantees}, as shown in Eq. \ref{eq:dynamicDP} below:
\begin{equation}
\mathcal{F}(x)=f_2\left(f_1(x)+\mathcal{N}(\sigma^2)\right),
\label{eq:dynamicDP}
\end{equation}
\noindent where it is assumed that the local model is a composition function of $f=f_2\circ f_1$. Note that when $x$ is vector-based, noises are sampled and added element-wise. A detailed sample algorithm can be found in \cite{CollaborativeInference22Ryu}.

\section{Multi-client SL with Different Privacy Requirements}\label{variousDP}
Considering the widespread data distribution, clients stand to benefit from participating in collaborative learning. In many practical scenarios, clients may have varying degrees of sensitivity, meaning they have different privacy expectations regarding their raw data. For example, some medical diseases (e.g., HIV or cancer) are more sensitive than others (e.g., fever or headache), and therefore, they require stronger privacy guarantees. However, the existing literature does not address this scenario in the context of SL. This section presents the first investigation of multi-client SL with different privacy requirements. The investigation includes identifying accuracy degradation and proposing an approach to enhance training performance.

\subsection{Different privacy requirements}
We initiate our preliminary experiments by configuring a setup involving $5$ clients and $1$ server. The training data is randomly and evenly distributed among the clients, adhering to the IID (independent and identically distributed) principle. In this context, most of the clients follow conventional SL training, while one or two clients require additional privacy guarantees with a budget of $\epsilon=2$ (along with a small value for $\delta$). These experiments utilize the foundational LeNet-5 model on the straightforward MNIST dataset \cite{LeNet98Lecun}. The learning model is split after the first pooling layer to distribute the head part to clients and the rest to the server, respectively.

\Rev{The clients ensure $(\epsilon,\delta)$-DP by adding Gaussian noise to their smashed data, as outlined in Eq. \ref{eq:Gaussian}, before transmitting it to the server.} Within each training round, the order of client training with the server is randomly shuffled. The post-training accuracy results are summarized in Table \ref{tab:mnist_acc_2dp}, revealing that incorporating the DP noise of some clients leads to a decrease in training accuracy for all clients compared to scenarios where all clients demand the same DP noise. In particular, when no client requires DP, the accuracy is $99.2\%$, which is slightly reduced to $98.8\%$ in the scenario of only one client is requiring DP. Regarding the DP-enabled client, its accuracy is $66.3\%$, which is significantly impacted compared to the averaged value of $73.4\%$ when all clients require the same DP noise. 
\Upd{The uniform privacy requirement, which allocates an equal privacy budget to each client, optimally facilitates achieving the highest accuracy regarding their respective noise levels. Thus, we consider the learning outcomes under uniform privacy as the benchmark upper limit for assessing accuracy loss or improvement in the context of varying privacy requirements.}
From the results, it is observed that DP-enabled clients are significantly affected while others are less so, which can lead to reluctance in contributing data from those clients with high privacy requirements. The experimental observations offer insights into what could be termed the `forgetting phenomenon' \cite{Forgetting99French} of SL, in which the server's learning process on smashed data is sequentially impacted by different distributions (smashed data with and without additional DP noise). In situations where the majority of clients require no noise, the server might `forget' the noise distribution while learning with `clean' smashed data. This ultimately diminishes accuracy when working with noisy smashed data from DP-enabled clients. Therefore, in the subsequent section, a novel training methodology is proposed to address this challenge.

\begin{table}[t]
\caption{Training accuracy of $5$ clients with varying numbers requiring DP noise ($\epsilon=2$) using conventional SL.}
\centering
\begin{tabular}{|c|c|c|c|c|c|}\cline{1-6}
\textbf{No. of clients} & \multicolumn{5}{c|}{\textbf{Training accuracy} ($\%$)} \\\cline{2-6}
\textbf{require noise} & $C_1$ & $C_2$ & $C_3$ & $C_4$ & $C_5$ \\\cline{1-6}
\textit{$0/5$ client} & \multicolumn{5}{c|}{$99.2$} \\\hline\cline{2-2}
\textit{$1/5$ client} & \multicolumn{1}{|c|}{$66.3$} & \multicolumn{4}{|c|}{$98.8$} \\\hline\cline{3-3}
\textit{$2/5$ clients} & \multicolumn{1}{|c|}{$70.1$} & \multicolumn{1}{c|}{$69.3$} & \multicolumn{3}{|c|}{$98.3$} \\\hline\cline{4-6}
\textit{$5/5$ clients} & \multicolumn{1}{|c|}{$74.0$} & $73.4$ & $72.9$ & $73.4$ & \multicolumn{1}{c|}{$73.5$} \\\hline\cline{2-6}
\end{tabular}
\label{tab:mnist_acc_2dp}
\end{table}

\subsection{Enhancing learning accuracy}
To mitigate the accuracy deterioration arising from training with different privacy requirements, we propose adapting server-side training, which involves the additional processing of smashed data to effectively address the forgetting phenomenon of multi-client SL.

\subsubsection{Reviewing other noise distributions}
Building upon the insights derived from the phenomenon of forgetting during sequential training with different noisy smashed data, we propose a strategy to address this by reviewing the noise distributions of clients with higher privacy requirements at the server side. Our approach involves introducing redundant data with increased noise levels for server-side training. This augmentation enhances the server's learning process, allowing it to review a more challenging noise distribution while learning from less noisy (or noise-free) smashed data. This helps alleviate the potential catastrophic forgetting that can arise during the sequential training of the server model across clients. To implement this proposed approach, we modify the execution of SL with multiple clients, as outlined in Alg. \ref{alg:multiClient}, with a particular focus on the server's role. \Upd{Building upon the foundational learning procedure of SL with multiple clients detailed in \cite{SplitLearning18Gupta},} our modification involves the server's preparation of supplementary data using incoming smashed data from clients (Alg. \ref{alg:multiClient}, line $6$). The specific steps for data preparation are illustrated in Alg. \ref{alg:prepareData}.

\begin{algorithm}[t]
\caption{One global epoch of SL.}
\begin{algorithmic}
\State $Clients$ and $Server$ receive their model parts and initialize corresponding weights
\end{algorithmic}
\begin{algorithmic}[1]
\State $Server$ sets $Client_0$ as the last trained client
\For{each $Client_i$ among all the $Clients$}
\State $Client_i$ requests weights from the last trained client \\ and updates its weights
\While{$Client_i$ has data to train with $Server$}
\State $Client_i$ performs forward propagation on its data \\ and sends output with labels to $Server$
\BeginBox
\State $Server$ prepares data using incoming features
\State $Server$ propagates the prepared data through its model and computes gradients for its output layer
\State $Server$ back-propagates the error to its first layer
\State $Server$ sends split layer's gradients to $Client_i$
\EndBox
\State $Client_i$ back-propagates the received gradients
\State $Client_i$ and $Server$ update their model weights
\EndWhile
\State $Server$ sets $Client_i$ as the last trained client
\EndFor
\end{algorithmic}
\label{alg:multiClient}
\end{algorithm}

\begin{algorithm}[t]
\caption{Data preparation at server.}
\begin{algorithmic}
\State \textbf{Input:} Incoming features with labels
\State \textbf{Output:} Prepared data for server training
\end{algorithmic}
\begin{algorithmic}[1]
\State $Server$ duplicates $Client_i$'s smashed data and labels
\State $Server$ adds additional noise to the duplicated smashed data and concatenates it with $Client_i$'s smashed data
\State $Server$ concatenates the duplicated labels with $Client_i$'s labels
\end{algorithmic}
\label{alg:prepareData}
\end{algorithm}

\Upd{In each iteration of client $C_i$'s training, upon receiving the smashed data (features), denoted as $z^t\triangleq f_{u_i}(x^t)$, and the corresponding labels, $y^t$, the server initiates by duplicating the data $(\tilde{z}^t,\tilde{y}^t)$ as depicted in Alg. \ref{alg:prepareData}, line $1$. Subsequently, it applies additional noise, $\mathcal{N}(\hat{\sigma})$, to the duplicated smashed data before concatenating to form $([z^t,\tilde{z}^t+\mathcal{N}(\hat{\sigma})],[y^t,\tilde{y}^t])$ as illustrated in lines $2$--$3$ of Alg. \ref{alg:prepareData}. For the vector-based smashed data, $z^t$, the additional noises are sampled and added element-wise.} The server then continues the propagation through its model part using the prepared data. Denoting the loss function that calculates the difference between the ground-truth labels and model-predicted outputs as $\mathcal{L}$, the computation of gradients at the server's last layer proceeds as follows:

\begin{equation}
\nabla\mathcal{L}(output,labels)=\underset{u_i,w}{\nabla}\mathcal{L}\Big(g_w\big([z^t,\tilde{z}^t+\mathcal{N}(\hat{\sigma})]\big),[y^t,\tilde{y}^t]\Big)
\label{eq:lossNoiseAdded}
\end{equation}

Using SGD, these gradients are backpropagated to update the weights of the server model part, starting from the last layer and moving towards the first layer. It is crucial to highlight that the computed gradients for the split layer have dimensions matching the concatenated data, rather than the size of the original smashed data, $z^t$. Therefore, the server must slice these gradients to align with the size of $z^t$ before transmitting them to the client (see line $9$ of Alg. \ref{alg:multiClient}) to complete the backpropagation. Referring back to Eq. \ref{eq:lossNoiseAdded}, these gradients are calculated by considering not only errors stemming from training with $C_i$'s smashed data but also errors originating from the data with augmented noise. This approach enables the server, by updating $g_w$ using these gradients, to simultaneously learn from the distribution of $C_i$'s smashed data and the distribution with additional noise.

Assuming that DP-enabled clients utilize the Gaussian mechanism ($\mathcal{N}(\sigma)$) to safeguard their data privacy, and each employing distinct privacy budgets (requirements), our objective is to improve training accuracy by empowering the server to review the distribution of the more challenging added DP noise (characterized by a larger noise scale) during its training. Since the sum of two normal distributions follows a normal distribution \cite{SumGuassians}, we formulate the scale of the augmented noise using the following equation:

\begin{equation}
\hat{\sigma}^2=\sigma_j^2-\sigma_i^2,
\end{equation}

\noindent where $\sigma_i$ and $\sigma_j$ denote the noise scales of the current client and the client with higher privacy requirements, respectively. We also make the assumption that the current client applies noise to its split layer (i.e., max-pool layer in our experiments). However, if noise is applied to the output of the preceding activation function prior to the max-pool, the equation remains applicable due to the inherent properties of operations such as max-pool, which preserve the additive noise.

\subsubsection{Performance analysis and evaluation}
Continuing from the scenario outlined in the preliminary experiments presented earlier involving $5$ clients with some requiring noise ($\epsilon=2$) to enhance their data privacy, we proceed to implement our proposed approaches aimed at modifying the server's training procedure to improve the accuracy of the severely impacted clients. The results, as depicted in Table \ref{tab:mnist_acc_2dp_new}, clearly demonstrate the efficacy of our proposed approach. Specifically, in scenarios where only one client requires noise, the accuracy improves from $66.3\%$ to $72.8\%$, while the accuracy of clients that require no noise remains at $98.5\%$. Similarly, in other settings, the accuracy of clients requiring noise ranges between $73-74\%$, closely aligning with the accuracy observed in the scenario where all clients demand the same level of noise. In additional experiments involving $10$ clients aimed at increasing the number of clients not requiring noise, we observe that the accuracy of the sole client requiring noise decrease to $62.9\%$. However, this accuracy is subsequently improved to $71.1\%$ through the application of our noise review approach. We also observe similar accuracy loss and improvement in experiments with an equal ratio of DP-enabled clients over the total number of clients (i.e., $1$ over $5$ and $2$ over $10$).

\begin{table}[t]
\caption{Training accuracy of clients with some requiring DP noise ($\epsilon=2$) using our proposed approach.}
\centering
\begin{tabular}{|c|c|c|}\cline{2-3}
\multicolumn{1}{c|}{} & \multicolumn{2}{c|}{\textbf{Training accuracy} ($\%$) \textbf{of}} \\\hline
\textbf{Ratio of clients} & \textbf{\textit{Client(s) with}} & \textbf{\textit{Clients without}} \\
\textbf{requiring noise} & \textbf{\textit{DP noise}} & \textbf{\textit{DP noise}} \\\hline
$1$ over $5$ & $72.8$ & $98.5$ \\\hline
$2$ over $5$ & $73.4$ & $98.0$ \\\hline\hline
$1$ over $10$ & $71.1$ & $98.6$ \\\hline
$2$ over $10$ & $72.3$ & $98.5$ \\\hline
\end{tabular}
\label{tab:mnist_acc_2dp_new}
\end{table}

In summary, in the context where only a minority of clients require enhanced privacy through DP, and the majority do not, our approach effectively elevates the accuracy of the significantly impacted clients to a level comparable to the benchmark, which represents the highest achievable accuracy under ideal uniform privacy requirements. The fewer the DP-enabled clients (over the total number of clients), the more accuracy loss and the more improvement using our proposed approach.

In the proposed approaches, we make modifications exclusively to the server's training procedure, while ensuring that the costs incurred at the clients remain unaffected. The additional expenses introduced at the server, such as storing duplicated data and performing concatenation, are considered manageable since the server is assumed to possess sufficient computational resources to accommodate multiple clients. Regarding privacy, the proposed approach aligns with privacy concerns as the server processes noisy smashed data that retains privacy guarantees due to the post-processing property of DP. As a result, the new training method neither escalates costs for the clients nor compromises the privacy of local data. However, there is an increased demand for computational and storage resources at the server, which is more feasible to address compared to the limitations of less powerful devices on the client side. Comprehensive evaluation of the proposed approach is presented in the evaluation section.

\section{Utility and Privacy Trade-offs in SL with DP}\label{trade-off}
\Edt{While neural network architectures vary widely across applications, they generally consist of conceptually simple units called neurons. Each neuron computes a weighted sum of its inputs along with a bias term, $\mathbf{z}=\mathbf{w}\cdot\mathbf{x}+b$, followed by an element-wise nonlinear transformation, $\mathbf{a}=\phi(\mathbf{z})$ \cite{WeightNormalization16Salimans}. Due to the post-processing property of DP, privacy is still guaranteed whether noise is introduced to the layer's output, $\mathbf{a}+\mathcal{N}$, or to the input of the layer, $\phi\left(\mathbf{w}\cdot(\mathbf{x}+\mathcal{N})+b\right)$. However, noise applied to the input data of a basic neuron block can be amplified due to the weight (e.g., $\mathbf{w}\cdot\mathcal{N}$) or the non-linearity (e.g., $\phi(\mathbf{z}+\mathbf{w}\cdot\mathcal{N})$), potentially having a higher impact on model utility. Drawing inspiration from \cite{AttackingProtecting22He}, which investigates the effects of adding noise to input or split layers to mitigate data leakage from reconstruction attacks, we empirically analyze the application of noise across different local layers of SL in this section. \Rev{Currently, there is a lack of theoretical analysis on the accuracy-privacy trade-offs when introducing noise across local layers in SL. Our empirical investigation seeks to underscore the potential for such analysis.} We begin by defining the threat model for our data leakage analysis, and then present by our findings on the optimal position for introducing noise to achieve the most favorable trade-offs between privacy and accuracy.}

\subsection{Threat model}
We consider SL with two parties, where multiple clients collaborate in learning with a server. The target model is split into two parts: $h_\theta=g_w\circ f_u$. The clients, performing $f_u$, are implicitly trusted to never leak their raw data to other parties. However, the server, performing $g_w$, is considered semi-trusted (honest-but-curious \cite{HonestButCurious14Paverd}), meaning it faithfully adheres to the training procedure but may have a certain level of curiosity about clients' private data. During collaborative learning, given smashed data, the server employs model inversion attacks \cite{ModelInversion19He,AttackingProtecting22He} to reveal clients' raw data. For simplicity and without loss of generality, we use a two-participant system (a client and a server) to evaluate the leakage of client's data.

Unlike the adversarial approach in \cite{AdversarialApproachAnalysis22Gursoy}, which analyzes DP mechanisms on categorical data, these model inversion attacks provide a means for assessing leakage on general data with high-quality visual results. Our investigation focuses on the attack in the black-box setting, known for its heightened effectiveness \cite{ModelInversion19He}. To facilitate black-box setting attacks, we assume that the server has the capability to query one of the clients to construct the inverse network, aligning with the assumption in \cite{ModelInversion19He}. By maximizing the adversary's capabilities and effectiveness, we aim to provide a more realistic reflection of DP's practical utility in safeguarding user private data.

In addition to visualizing reconstructed images, various metrics such as \Upd{Mean Squared Error} (MSE), \Upd{Structural Similarity Index Measure} (SSIM), \Upd{Peak
Signal-to-Noise Ratio} (PSNR) \cite{ImageQuality04Zhou} and Kullback-Leibler divergence have been employed in the literature \cite{SLon1DCNN20Sharif,ModelInversion19He,DroppingActivation18Dong,NoPeek20Vepakomma} to quantify the effectiveness of the reconstruction. These metrics typically gauge the disparity between the original images and their reconstructed counterparts. In our study, we primarily employ SSIM \cite{SSIM04Wang} as the key metric for assessing the similarity between raw and reconstructed data, which effectively quantifies data leakage. SSIM, being a perceptual metric, evaluates the degradation of image quality, offering a more intuitive and interpretable measure compared to other alternatives \cite{CollaborativeInference22Ryu}.

\subsection{Trade-offs between accuracy and privacy budget}

We conduct preliminary experiments using the LeNet-5 model on the MNIST dataset. In the context of SL, we explore two potential model splitting points after each pooling layer, denoted as Split-1 and Split-2, for distribution between clients and the server. Table \ref{tab:cutPoints} provides an overview of the model parts allocated to clients and the server, corresponding to the splitting points after the first and second convolutional groups. A convolutional group consists of a 2D-convolution layer (\textsf{Conv}), followed by an activation function and, if applicable, a pooling operation. We use the rectified linear unit (\textsf{ReLU}) for activation and the max-pool operation (\textsf{MaxP}) for pooling. As a result, layers such as \textsf{Conv(1)}, \textsf{ReLU(1)}, and \textsf{MaxP(1)} belong to the first convolutional group (1st Conv.), while \textsf{Conv(2)}, \textsf{ReLU(2)}, and \textsf{MaxP(2)} constitute the second convolutional group (2nd Conv.). These configurations are tested in a single-client setting, and with these settings, we achieve an accuracy exceeding $99\%$, which is comparable to the current state-of-the-art results \cite{MNISTAccuracy}.

\begin{table}[t]
\centering
\caption{Distribution of model parts between the client and server.}
\begin{tabular}{|c|c|c|}\hline
\textbf{Case} & \textbf{Client model part} & \textbf{Server model part} \\\hline
\textbf{\textit{Split-1}} & 1st Conv. \textcolor{white}{-- 2nd Conv.} & 2nd Conv. -- 1st Dense -- 2nd Dense \\\hline
\textbf{\textit{Split-2}} & 1st Conv. -- 2nd Conv. & \textcolor{white}{2nd Conv. --} 1st Dense -- 2nd Dense \\\hline
\end{tabular}
\label{tab:cutPoints}
\end{table}

We introduce Gaussian noise (Eq. \ref{eq:Gaussian}) to every local layer and evaluate its impact on the model's accuracy.
\Upd{Additionally, we clamp the layer's output between $[0,1]$ to acquire $1$-sensitive outcomes and systematically vary the $\epsilon$ values.}
Subsequently, we plot the accuracy of the affected model against the corresponding privacy budgets to analyze the inherent trade-offs.

\begin{figure}[h]
\centering
\subfloat[Within the 1st Conv., Split-1]{
\includegraphics[width=.23\textwidth]{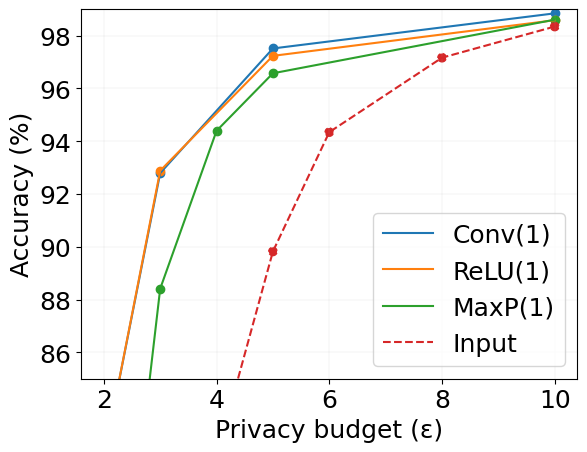}
\label{fig:mnist_acc_eps_1st}
}
\hfill
\subfloat[Within the 2nd Conv., Split-2]{
\includegraphics[width=.23\textwidth]{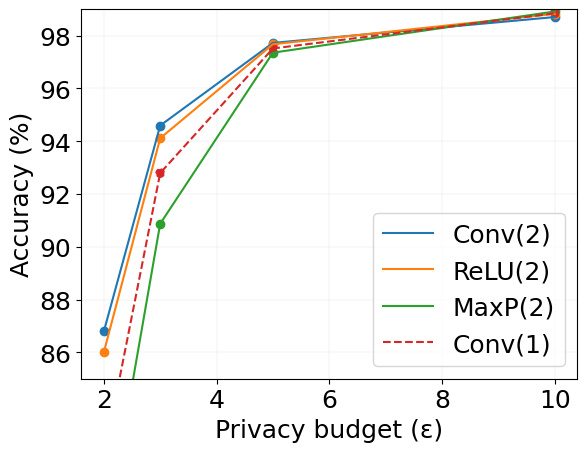}
\label{fig:mnist_acc_eps_2nd}
}
\caption{Trade-offs between accuracy and privacy budgets, with LeNet-5 on MNIST.}
\label{fig:mnist_acc_eps}
\end{figure}

The results align with our expectations, showing a decrease in accuracy with a reduction in the privacy budget, $\epsilon$. Based on the collected results visualized in Fig. \ref{fig:mnist_acc_eps}, it becomes evident that applying DP noise to the input layer results in the most significant accuracy degradation (Fig. \ref{fig:mnist_acc_eps_1st}, the dashed line). In contrast, introducing noise to the \textsf{Conv} and \textsf{ReLU} layers produces comparable results. However, a slight accuracy reduction is observed when noise is introduced to the pooling layer, which can be attributed to the data size reduction caused by the \textsf{MaxP} operation (the green lines). Moreover, applying noise to the deeper layers, specifically the 2nd Conv. (Fig. \ref{fig:mnist_acc_eps_2nd}, the solid lines), results in improved accuracy compared to the shallower layers, namely the 1st Conv. (Fig. \ref{fig:mnist_acc_eps_2nd}, the dashed line). This improvement can be attributed to the reduced amplification of noise within those layers.

\subsection{Trade-offs between accuracy and data leakage}
While noise-based mechanisms theoretically guarantee DP within a specified budget, the nature of model splitting introduces a potential vulnerability, allowing raw data leakage from the smashed data. This vulnerability poses a challenge to the privacy principles inherent in the design of SL. To quantitatively assess the privacy loss attributed to data leakage, we evaluate the quality of reconstructed data (using model inversion attacks) by comparing it to raw data, utilizing the SSIM metric. For illustrative purposes, we plot accuracy against dissimilarity results, formulated as (1--SSIM), to facilitate the observation of trade-offs.
\Upd{Note that the SSIM values range between $[0,1]$, with $1$ indicating perfect similarity and $0$ representing no similarity. Therefore, larger dissimilarity values suggest better privacy preservation as there is less leakage from the reconstructed data.}

Continuing our preliminary experiments, we implement the model inversion attack on the server, adhering to a black-box setting, as described in the threat model. We then assess the effectiveness of the attack when DP noise is introduced to different local layers, with the deep model partitioned as per the configuration specified earlier in Table \ref{tab:cutPoints}.

Similar to evaluating the trade-offs between accuracy and privacy budget, we observe analogous trade-offs when introducing noise to different layers in the same convolutional group. Notably, introducing noise to the input layer significantly degrades accuracy, accompanied by a substantial reduction in reconstruction quality. This pattern results in trade-offs comparable to those witnessed in other layers within the 1st Conv.

\begin{figure}[h]
\centering
\subfloat[Within the 1st Conv., Split-1]{
\includegraphics[width=.23\textwidth]{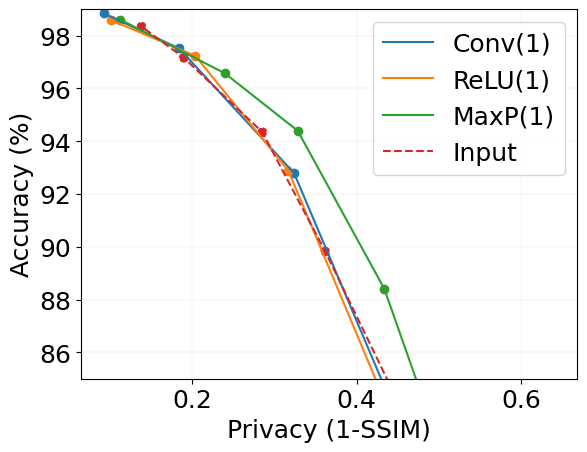}
}
\hfill
\subfloat[Within the 2nd Conv., Split-2]{
\includegraphics[width=.23\textwidth]{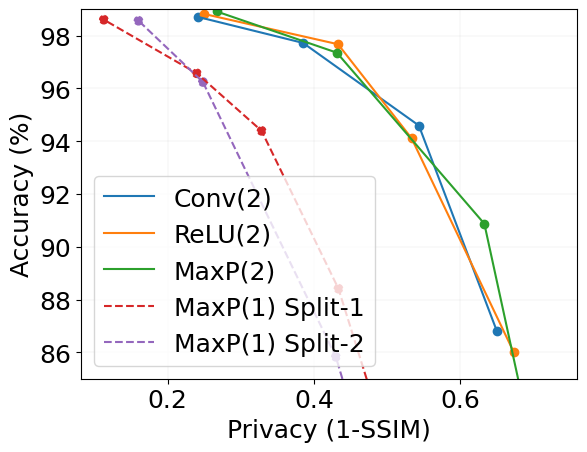}
\label{fig:mnist_acc_sim_2nd}
}
\caption{Trade-offs between accuracy and data leakage, with LeNet-5 on MNIST.}
\label{fig:mnist_acc_sim}
\end{figure}

As depicted in Fig. \ref{fig:mnist_acc_sim}, the most favorable trade-offs are achieved by adding noise to the layers within the 2nd Conv. While adding noise to the 2nd Conv. helps improve accuracy slightly (see Fig. \ref{fig:mnist_acc_eps_2nd}), it significantly impacts the reconstruction results, leading to superior privacy compared to noise in the 1st Conv. Particularly, with a similar amount of sacrificed accuracy, adding noise to the 2nd Conv. results in double the privacy compared to the 1st Conv. This observation aligns with the findings from \cite{SLon1DCNN20Sharif}, where the authors claim that adding more hidden layers on the client side helps further reduce data leakage from visualizing smashed data. Even though leakage from visualizing smashed data is not as significant as leakage from model inversion attacks \cite{BinarizingSL23Pham}, adding more hidden layers (convolutional groups) also helps further reduce leakage from model inversion attacks, hence strengthening privacy preservation. However, these results lead to another trade-off related to computational costs on the client side because more layers significantly increase the local overhead of the clients, which are typically low-end devices with limited resource capabilities.

Another observation is that within a convolutional group, introducing noise to \textsf{Conv}, \textsf{ReLU}, or \textsf{MaxP} layers results in similar trade-offs, with the most favorable trade-off occurring at the pooling layers. In Fig. \ref{fig:mnist_acc_sim_2nd}, we also plot the trade-off when noise is applied to the \textsf{MaxP(1)} layer, and we attempt to reconstruct the raw data from the smashed data obtained in Split-1 and Split-2. Surprisingly, even though noise propagates through the 2nd Conv. in Split-2, the trade-off is comparable to that in Split-1, which confirms the insight into achieving a better trade-off when noise is introduced in the latter layers. Hence, this finding explains the common practice of introducing noise to the split layer in much of the literature. 

In summary, based on our assessment through these preliminary experiments, it becomes evident that applying noise to deeper layers (preferably closer to the split layer) yields more favorable trade-offs between accuracy and privacy. Typically, in SL, the sub-sampling or pooling layer is chosen as the split layer to reduce client-server communication overhead. Taking the privacy budget into account, the application of DP noise to the layer immediately preceding the pooling (split layer) yields the most optimal trade-off. When considering data leakage, applying DP noise to the split layer yields the best balance between accuracy and privacy trade-offs. Regarding introducing noise to the input layer, which leads to high privacy preservation but also significantly impacts accuracy, we will discuss and propose an approach to improve accuracy while aiming for a more balanced trade-off.

\subsection{Enhancing performance of SL-DP} 
Regarding privacy preservation, introducing noise to the input layer before sending it directly to the server offers noticeable advantages. For example, exposing smashed data is more minor, resulting in decreased data leakage (as demonstrated earlier) and lower communication costs. However, this approach comes at the expense of a significant impact on model accuracy, which remains a primary concern. Conversely, applying noise to deeper layers, such as the 1st or 2nd Conv., results in higher model accuracy and more favorable trade-offs. Nevertheless, the smashed data generated from these subsequent layers is more extensive, increasing communication costs for training and inference phases.
\Upd{Building upon these insights and capitalizing on the introduction of noise at the split layer, we propose a strategy to diminish the size of smashed data, thereby reducing data leakage, akin to introducing noise to the input layer. Concurrently, our aim is to decrease the communication costs of SL by transmitting smaller data volumes between clients and the server.}

Consider a deep model, $h=g\circ \mathcal{F}$, divided into two parts, $\mathcal{F}$ and $g$, distributed to the clients and server, respectively. Here, $\mathcal{F}$ is the differentially private version of $f$, denoted as $\mathcal{F}=f+\mathcal{N}(\sigma)$. We propose appending a small component with a few layers, denoted as $f'$, which converts the smashed data to the size of the input. The new deep model is then formulated as:
\begin{equation}
\tilde{h}=(g\circ f)\circ(f'\circ\mathcal{F}),
\label{eq:upSampling}
\end{equation}
\noindent where $f'\circ\mathcal{F}$ is distributed to the clients, while $g\circ f$ remains on the server. In this proposed deep model design, we increase the computational cost at the client due to the processing of $f'$, while the server incurs additional costs for executing $f$. Fortunately, these increased costs at the server are manageable due to its resource assumptions, while the additional costs at the client can be considered a trade-off for bandwidth savings. Additionally, smaller sizes of exposed data contribute to limiting potential data leakage. Considering the selected LeNet-5 model, which is based on a convolutional neural network (CNN), we design $f'$ as a transposed convolutional (de-convolutional) layer \cite{Deconvolutional11Zeiler} for up-sampling the smashed data to match the original input data size. Given the widespread use of CNN-based deep models across various domains and data types \cite{SurveyCNN20Khan}, this design is widely applicable. We will conduct a performance evaluation and analyze the accuracy-privacy trade-offs of our proposed approach in the following section.


\section{Experimental Evaluation}\label{experiment}
This section discusses the comprehensive experiments we conducted using a deeper model and a more challenging dataset to thoroughly assess the trade-offs between accuracy and privacy and evaluate the proposed approaches.

We select CIFAR-10 \cite{CIFAR09Krizhevsky}, an established computer vision dataset that is a subset of the $80$ million tiny images dataset. CIFAR-10 comprises $60k$ $32\times32$ color images distributed across $10$ distinct object classes, with $6k$ images per class. For our experiments, we utilize a VGG-11 \cite{VGG15Simonyan} based deep model. We split the model after the first pooling layer, with the client handling the head part and the server managing the remaining layers. The architectural details of the deep model, including layer structure and splitting point, are presented in Table \ref{tab:vggModel}.

\begin{table}[ht]\centering
\caption{Architecture of the selected VGG-11-based deep model.}
\begin{tabular}{|r|p{.7\linewidth}|}\hline
\textbf{Client part} & Input -- \textsf{C3-32} -- \textsf{C3-32} -- \textsf{MaxP} \ \ \ \ $<$--\textit{Split}--$>$  \\\hline
\textbf{Server part} & \textsf{C3-64} -- \textsf{C3-64} -- \textsf{MaxP} -- \textsf{C3-128} -- \textsf{C3--128} -- \textsf{MaxP} -- \textsf{C3-256} -- \textsf{C3-256} -- \textsf{MaxP} -- \textsf{D128} -- \textsf{D10} -- Output \\\hline
\end{tabular}
\label{tab:vggModel}
\end{table}

In Table \ref{tab:vggModel}, the notation \textsf{C3-32} denotes a convolution layer with $32$ filters of size $3$, followed by a \textsf{ReLU} activation function. Similarly, \textsf{D128} represents a fully connected (dense) layer with $128$ output units. Following the naming convention used in our preliminary experiments, we label the layers in the client's model part as follows in our plots: Input, \textsf{Conv(1)}, \textsf{ReLU(1)} (within the 1st Conv.), \textsf{Conv(2)}, \textsf{ReLU(2)}, and \textsf{MaxP} (within the 2nd Conv.). Throughout our experiments using PyTorch on Google Colab, we conduct model training for $100$ epochs, and employing the Adam optimizer \cite{Adam17Kingma} with default parameters and a batch size of $64$. The learning rate is adjusted using a cosine annealing scheduler \cite{CosineAnnealingLR}, with the maximum number of iterations set equal to the number of epochs. After training, we evaluate the model (by combining the client's part with the server's model part) on the test set and present the results accordingly.

\subsection{Training with different noise levels}
In our preliminary experiments, we showcase the effectiveness of the proposed approach, reviewing other noise distributions, in enhancing the accuracy of clients requiring DP noise in a context characterized by different privacy requirements. As the number of DP-enabled clients reduces, the impact on accuracy under conventional SL becomes more pronounced, and our approach offers significant potential for accuracy improvement. Here, we delve deeper into evaluating the proposed approach, specifically focusing on its effectiveness across a broader range of noise levels and a more diverse set of privacy requirements.

We configure a setup with $10$ clients, each receiving a similar amount of data from all classes in the training set. However, only some of the clients introduce distinct levels of DP noise to the smashed data before transmitting it to the server for training, while the remaining clients do not require any noise. We set $3$ clients require noise with the privacy budgets are varied as follows: $\{2,3,4\}$ for experiments with the MNIST dataset and $\{2,3,5\}$ for the CIFAR10 dataset. Following the training phase, we assess the accuracy of each client and report the results for both SL with and without our proposed approach. Additionally, we establish a benchmark by considering the results with conventional SL when all clients have the same privacy budget. For instance, when all clients have a privacy budget of $\epsilon=2$, the averaged accuracy serves as a reference point for comparison. This benchmark represents the highest accuracy achievable by a client with a specific required noise level in an ideal environment with uniform privacy requirements. 

\begin{table}[b]
\caption{Training accuracy ($\%$) of $10$ clients with different required noise levels on the MNIST dataset.}
\centering
\begin{tabular}{|l|c|c|c|c|}\hline
\multicolumn{1}{|r|}{\textit{Client}} & $C_1$ & $C_2$ & $C_3$ & $C_4-C_{10}$ \\\hline
\textbf{\textit{Scheme}} & $\epsilon=2$ & $\epsilon=3$ & $\epsilon=4$ & \textit{No DP required} \\\hline
Benchmark & 74.3 & 88.4 & 94.1 & 99.1 \\\hline\hline
Conventional SL & 63.4 & 82.7 & 90.5 & 98.6 \\\hline
Our approach & 70.3 & 86.6 & 92.6 & 98.4 \\\hline
\multicolumn{1}{r|}{\textit{Improvement}} & \textit{$+$6.9} & \textit{$+$3.9} & \textit{$+$2.1} & \textit{$-$0.2} \\\cline{2-5}
\end{tabular}
\label{tab:iid_MNIST}
\end{table}

The results are presented in Tables \ref{tab:iid_MNIST} and \ref{tab:iid_CIFAR} for the experiments conducted on the MNIST and CIFAR10 datasets, respectively. Upon analyzing the MNIST results, it is apparent that after training with conventional SL, the accuracy of all clients decreases when compared to the benchmark. This decrease can be attributed to the forgetting phenomenon that arises during sequential learning on data with different added noise distributions. Clients requiring the highest noise experience the most significant drop in accuracy due to the influence of high-scale noise and the associated forgetting during the learning process. When applying our proposed approach, the accuracy of these high-noise clients shows a substantial improvement, approaching the benchmark. For clarity, we also provide the degree of improvement (our approach compared to conventional SL). Notably, the most significant improvement is observed with clients requiring higher levels of noise, with the improvement gradually decreasing for clients with lower noise requirements. It's worth noting that the accuracy of clients not requiring noise remains similar to that of conventional SL, with only a negligible $0.2\%$ difference. However, both show a slight reduction compared to the benchmark. The reduction in accuracy compared to the benchmark can be attributed to the server being trained on diverse data distributions, resulting in a larger and more complex dataset (comprising of noisy smashed data from all clients) for the server model to learn. 

\begin{table}[t]
\caption{Training accuracy ($\%$) of $10$ clients with different required noise levels on the CIFAR10 dataset.}
\centering
\begin{tabular}{|l|c|c|c|c|}\hline
\multicolumn{1}{|r|}{\textit{Client}} & $C_1$ & $C_2$ & $C_3$ & $C_4-C_{10}$ \\\hline
\textbf{\textit{Scheme}} & $\epsilon=2$ & $\epsilon=3$ & $\epsilon=5$ & \textit{No DP required} \\\hline
Benchmark & 69.3 & 76.9 & 82.4 & 88.4 \\\hline\hline
Conventional SL & 31.0 & 51.9 & 70.5 & 86.5 \\\hline
Our approach & 63.9 & 73.5 & 80.4 & 86.0 \\\hline
\multicolumn{1}{r|}{\textit{Improvement}} & \textit{$+$33} & \textit{$+$22} & \textit{$+$9.9} & \textit{$-$0.5} \\\cline{2-5}
\end{tabular}
\label{tab:iid_CIFAR}
\end{table}

Table \ref{tab:iid_CIFAR} presents the results of experiments with the CIFAR10 dataset. Because this dataset is more challenging, we observe a significant drop in accuracy when using conventional SL compared to the benchmark. The high level of noise not only damages the accuracy of clients requiring more noise but also impacts the accuracy of clients requiring less noise. However, when applying our approach, we can significantly improve the accuracy of all clients that require noise. For example, client $C_1$, which requires the most noise with $\epsilon=2$, experiences the highest accuracy drop of approximately $38\%$. However, our approach narrows this gap, achieving only a $5\%$ difference from the benchmark. These results reveal a consistent trend of improvement, with the most noise-required clients benefiting the most, while clients with lower noise requirements also experience gradual enhancements. Finally, the accuracy of clients not requiring noise decreases slightly (by $0.5\%$). This trade-off occurs because those clients review other clients' noisy data during training. However, this trade-off is justifiable given the significant improvement observed in clients requiring noise. These experiment results demonstrate the efficiency of our proposed approach, encouraging clients with more sensitive data to participate in collaborative training and achieve high learning performance.

\subsection{Trade-offs between accuracy and privacy}
We investigate the impact of applying DP noise to different local layers on model accuracy and data reconstruction results.

\begin{figure}[b]
\centering
\subfloat[Within the 1st Conv.]{
\includegraphics[width=.23\textwidth]{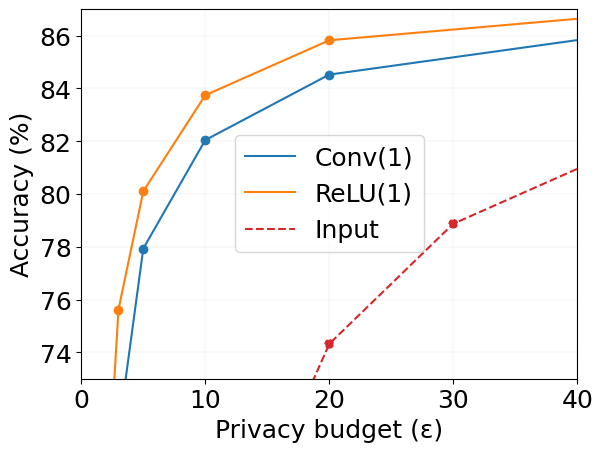}
}
\hfill
\subfloat[Within the 2nd Conv.]{
\includegraphics[width=.23\textwidth]{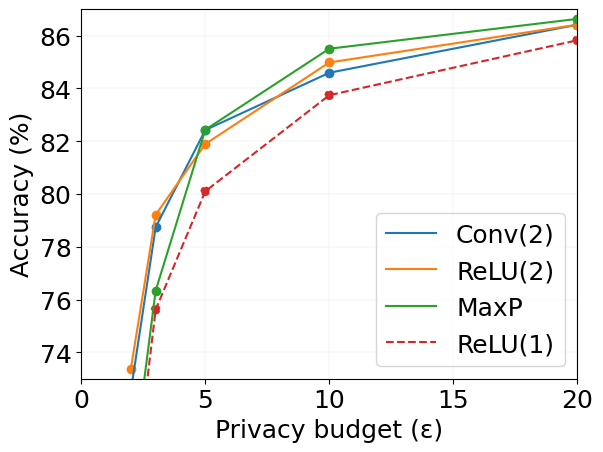}
\label{fig:cifar_acc_eps_2nd}
}
\caption{Trade-offs between accuracy and privacy budgets, with VGG-11 on CIFAR10.}
\label{fig:cifar_acc_eps}
\end{figure}

Fig. \ref{fig:cifar_acc_eps} plots the trade-offs between accuracy and privacy budgets, illustrating alignment with the findings from our previous preliminary experiments. When noise is added to the input layer, it has the most substantial impact on accuracy. Conversely, applying noise to the \textsf{Conv} or the subsequent activation layer, \textsf{ReLU}, yields comparable accuracy impacts. Notably, noise introduced in later layers (e.g., \textsf{Conv(2)}, \textsf{ReLU(2)}) results in less accuracy reduction, offering a more favorable trade-off. While adding DP noise to the split layer (\textsf{MaxP}) causes a minor decrease in accuracy compared to \textsf{Conv(2)} or \textsf{ReLU(2)}, we still recommend applying DP noise to the layer immediately preceding the split layer for achieving the optimal trade-off between accuracy and the privacy budget.

\begin{figure}[t]
\centering
\subfloat[Within the 1st Conv.]{
\includegraphics[width=.23\textwidth]{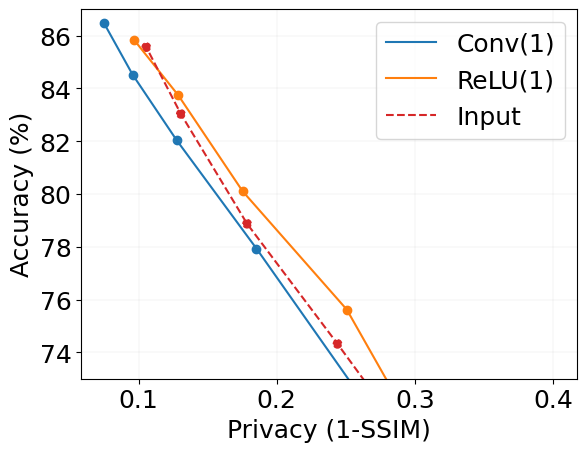}
}
\hfill
\subfloat[Within the 2nd Conv.]{
\includegraphics[width=.23\textwidth]{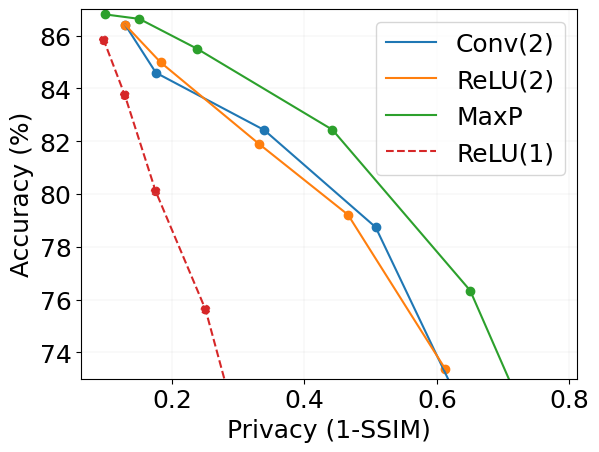}
\label{fig:cifar_acc_sim_2nd}
}
\caption{Trade-offs between accuracy and data leakage, with VGG-11 on CIFAR10.}
\label{fig:cifar_acc_sim}
\end{figure}

Fig. \ref{fig:cifar_acc_sim} illustrates the trade-offs between accuracy and privacy concerning data leakage. The findings from Fig. \ref{fig:cifar_acc_sim_2nd} are quite apparent: applying noise to latter layers leads to a more favorable trade-off, exhibiting worse reconstruction results while sacrificing similar levels of accuracy. Within a convolutional group, comparable trade-off are observed when noise is added to the \textsf{Conv} or the subsequent \textsf{ReLU} layer. Notably, applying noise to the \textsf{MaxP} layer yields the optimal trade-off due to sub-sampling data. Consequently, adhering to the conventional practice of applying noise to the split layer at the client side, as seen in the literature, indeed yields the best compromise between accuracy and data privacy.

The results with the CIFAR10 dataset align with the preliminary experiments, confirming our findings about the accuracy-privacy trade-off when introducing DP to SL. Based on these findings, privacy practitioners can select appropriate strategies to achieve the most favorable outcomes.

\begin{figure}[b]
\centering
\includegraphics[width=.24\textwidth]{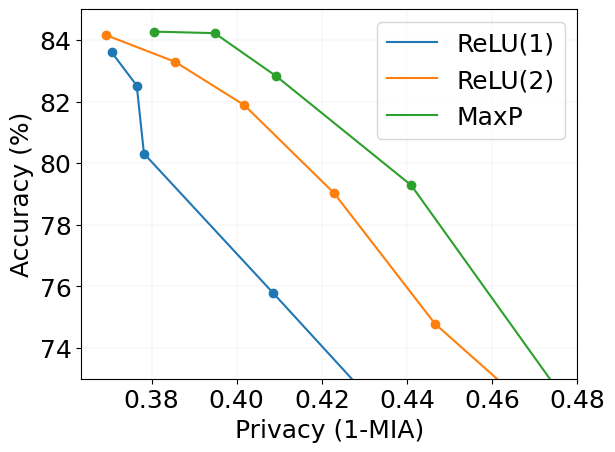}
\caption{Trade-offs between accuracy and privacy leakage from model inversion attacks.}
\label{fig:cifar_acc_mia}
\end{figure}

\Edt{Besides the information leakage from the smashed data due to the splitting manner of SL, privacy in ML/DL models is also impacted by other privacy attacks, such as membership inference attacks (MIAs) \cite{MembershipInference17Shokri}. We further investigate the impact of introducing DP noise on different SL local layers and its effect on the accuracy-privacy trade-offs, with privacy measured using MIA. Following the attack strategy outlined in \cite{PracticalMIA22Chen}, we assume collusion between a client and the server to achieve optimal attack performance. The local model of the colluding client acts as a shadow model, which combined with the server model, is used to train an attack model that targets any client on the client side. The goal is to verify whether the victim client contains samples from known datasets. The accuracy-privacy trade-offs are plotted in Fig. \ref{fig:cifar_acc_mia}, with privacy is measured by MIA results. Note that the y-axis shows the values of (1--MIA), where higher values indicate higher privacy (less leakage). The results exhibit a trend similar to model inversion attacks, where adding noise at a later layer leads to more favorable trade-offs between sacrificed accuracy and improved privacy preservation.}

\subsection{Enhancing performance of SL-DP}
We initially evaluate the performance of our proposed approach to enhance performance of SL-DP by continuing the preliminary experiments using LeNet-5 and the MNIST dataset.
\Upd{To implement the splitting strategies outlined in Eq. \ref{eq:upSampling}, we incorporate a transposed convolutional layer (ConvTranspose) to up-sample the smashed data to the original size of the input images. Consequently, the server model is now transformed into the full model ($g\circ f$) to handle the smashed data of the original input size. The network architectures for the two splits are provided in Table \ref{tab:cutPoint2}.}

\begin{table}[ht]
\centering
\caption{A proposed distribution of model parts between the client and server.}
\begin{tabular}{|l|l|l|}\hline
\textbf{Case} & \textbf{Client model part} & \textbf{Server model part} \\\hline
\textbf{\textit{Split-1}} & 1st Conv. -- ConvTranspose & LeNet-5 model \\\hline
\textbf{\textit{Split-2}} & 1st Conv. -- 2nd Conv. -- ConvTranspose & LeNet-5 model \\\hline
\end{tabular}
\label{tab:cutPoint2}
\end{table}

\begin{figure}[t]
\centering
\subfloat[Accuracy vs privacy budget]{
\includegraphics[width=.23\textwidth]{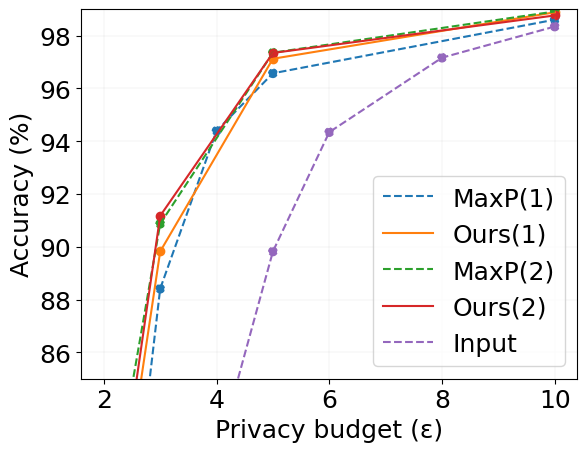}
\label{fig:mnist_acc_eps_our}
}
\hfill
\subfloat[Accuracy vs data leakage]{
\includegraphics[width=.23\textwidth]{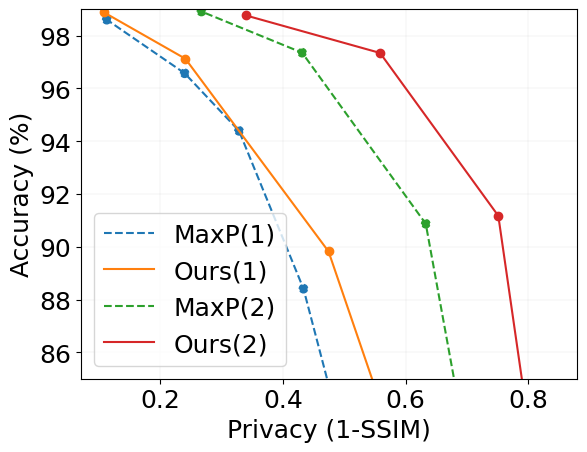}
\label{fig:mnist_acc_sim_our}
}
\caption{Accuracy and privacy trade-offs of our approach on the MNIST dataset.}
\label{fig:mnist_acc_our}
\end{figure}

Fig. \ref{fig:mnist_acc_eps_our} presents the trade-off between accuracy and the privacy budget, illustrating that our proposal maintains the comparable accuracy trend as when introducing noise to the \textsf{MaxP} layers in both Split-1 and Split-2. Using our proposed approach, the smashed data has the same size as the input data, significantly reducing transmission costs and data exposure. Fig. \ref{fig:mnist_acc_sim_our} illustrates the trade-offs between accuracy and data leakage, showing that our approach is more favorable than introducing noise to the \textsf{MaxP} layer, which achieve optimal trade-offs in our preliminary experiments.

\begin{figure}[t]
\centering
\subfloat[Accuracy vs privacy budget]{
\includegraphics[width=.23\textwidth]{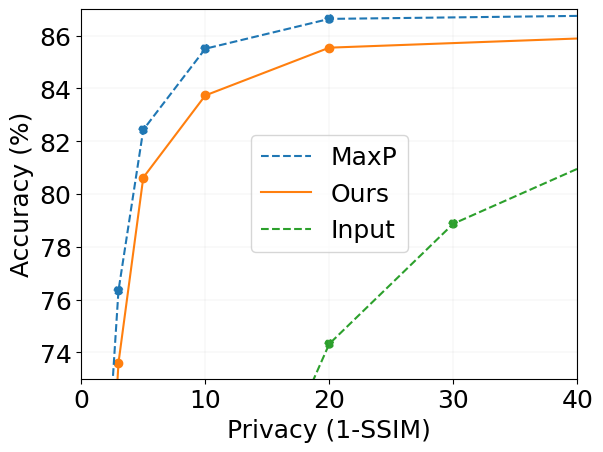}
\label{fig:cifar_acc_eps_our}
}
\hfill
\subfloat[Accuracy vs data leakage]{
\includegraphics[width=.23\textwidth]{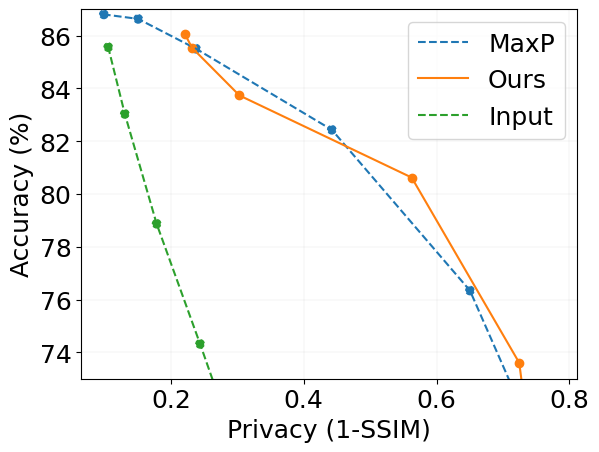}
\label{fig:cifar_acc_sim_our}
}
\caption{Accuracy and privacy trade-offs of our approach on the CIFAR10 dataset.}
\label{fig:cifar_acc_our}
\end{figure}

Further evaluation of the CIFAR10 dataset with a deeper model yields similar results, as shown in Fig. \ref{fig:cifar_acc_our}, where our approach maintains the accuracy of noise introduced at the split layer while achieving low leakage similar to noise introduced at the input layer. The result demonstrates that our approach provides the most favorable trade-offs compared to introducing noise to any SL local layers, with only a marginal additional computation overhead. Regarding the additional local computation cost, it is minimal since the appended layer is very thin, and we can offset the saving communication costs as a return. The amount of communication costs saved (bandwidth, energy, and time) depends on the dataset size and the number of training rounds (epochs). In our experiments, considering that an image from CIFAR10 has the size of $3\times32\times32$ (channels $\times$ width $\times$ height), the smashed data size (without our approach) is $32\times16\times32$, and the dataset's size is $50k$, with training conducted for $100$ epochs. Utilizing our proposed approach could save approximately $5.3\times$ communication costs during the training procedure, which is a significant advantage for low-end devices.

\subsection{Experiments on time-series data and discussions}
\Upd{In this section, we present additional experiments on 1D time-series data and discuss the limitations of the proposed approaches. Following the methodology of \cite{SLon1DCNN20Sharif}, we conduct experiments on an ECG dataset to further evaluate the efficiency of our proposed scheme. We set up a two-layer 1D CNN model to classify five heart diseases on the ECG dataset provided by the authors. All other configurations follow \cite{SLon1DCNN20Sharif}, except that we split and deploy the first convolutional layer to the clients. In the experiments, we distribute the dataset equally among $5$ clients who collaborate in training with a central server. We set up one client to require DP noise to provide stronger protection for its local data while others do not. We vary the privacy budget and report the results of training accuracy in Table \ref{tab:ecg_acc_dps_new}. The results exhibit similar trends observed in experiments with image datasets, validating the effectiveness of our approach in significantly improving the training accuracy of clients requiring DP noise compared to conventional SL.}

\Upd{Besides the advantages in significantly improving the accuracy of DP-required clients, from Tables \ref{tab:mnist_acc_2dp_new}, \ref{tab:iid_MNIST}, \ref{tab:iid_CIFAR}, and \ref{tab:ecg_acc_dps_new}, we can observe a slightly reduced accuracy of clients that do not require DP as a trade-off. However, the reduction is minor (within $0.5\%$) compared to the accuracy gained from the clients requiring DP. These benefits are attributed to the extra overheads at the server to process the training on noise-augmented data while the computation costs at the client side remain the same to conventional SL. Even though the server is normally assumed to have \emph{enough resources} to coordinate the training, the additional costs could lead to more resource consumption and impacted processing latency. In Table \ref{tab:mnist_acc_2dp_new}, we can also observe the effect of our approach in terms of scalability. In the case of $10$ clients, but with the same ratio of clients requiring noise i.e. $1$ over $5$ and $2$ over $10$, the accuracy results of both clients with and without noise are similar, demonstrating that our approach is stable with scalability.}

\begin{table}[t]
\caption{\Upd{Training accuracy of clients with one requiring DP noise on the ECG dataset.}}
\centering
\begin{tabular}{|c||c|c||c|c|}\cline{2-5}
\multicolumn{1}{c|}{} & \multicolumn{4}{c|}{\textbf{Training accuracy} ($\%$) \textbf{of client(s)}} \\\hline
\textbf{Privacy} & \textbf{\textit{With}} & \textbf{\textit{Without}} & \textbf{\textit{With}} & \textbf{\textit{Without}} \\
\textbf{budget} & \textbf{\textit{DP noise}} & \textbf{\textit{DP noise}} & \textbf{\textit{DP noise}} & \textbf{\textit{DP noise}} \\\hline
$\epsilon=3$ & $72.7$ & $97.7$ & $64.3$ & $98.1$ \\\hline
$\epsilon=5$ & $86.3$ & $98.1$ & $80.0$ & $98.5$ \\\hline
$\epsilon=7$ & $89.8$ & $98.1$ & $87.6$ & $98.5$ \\\hline
$\epsilon=10$ & $94.5$ & $98.5$ & $92.7$ & $98.6$ \\\hline
\multicolumn{1}{c|}{} & \multicolumn{2}{c|}{\textbf{\emph{Our approach}}} & \multicolumn{2}{c|}{\textbf{\emph{Conventional SL}}} \\\cline{2-5}
\end{tabular}
\label{tab:ecg_acc_dps_new}
\end{table}

\begin{figure}[h]
\centering
\includegraphics[width=.3\textwidth]{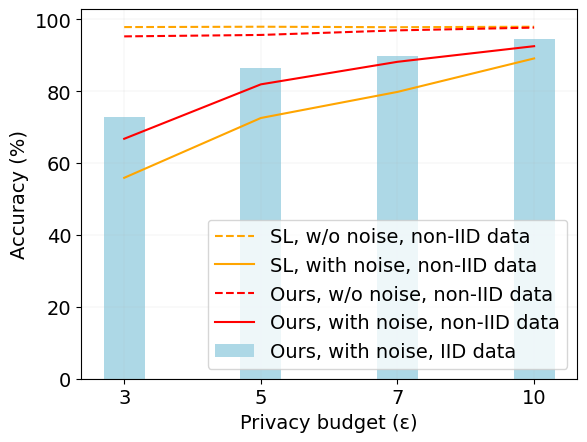}
\caption{Training performance on the ECG dataset with non-IID data distribution.}
\label{fig:ecg_acc_dps_non}
\end{figure}

\Edt{In earlier experiments, the training data is uniformly distributed among the clients. Next, we conduct experiments where data is distributed following non-IID settings to account for a more realistic scenario. Specifically, for the ECG dataset with five classes, the data is distributed to five clients such that each client $i$ receives $60\%$ of the samples from class $i$ and $40\%$ from other classes. From the experiment results shown in Fig. \ref{fig:ecg_acc_dps_non}, we can observe that the learning performance is further impacted by the data distribution (line vs bar charts). In this non-IID setting, the client requiring the highest level of DP noise experiences about a $5\%$ reduction in accuracy compared to the IID setting. However, we still observe an improvement trend when applying our proposed approach to conventional SL. While there is a performance gap between clients requiring noise and those that do not, our approach helps significantly reduce this gap with only a slight impact on the performance of clients that do not require noise. This is beneficial for collaborative learning scenarios with varying privacy demands.}

\begin{figure}[t]
\centering
\subfloat[Compared to SL-DP]{
\includegraphics[width=.23\textwidth]{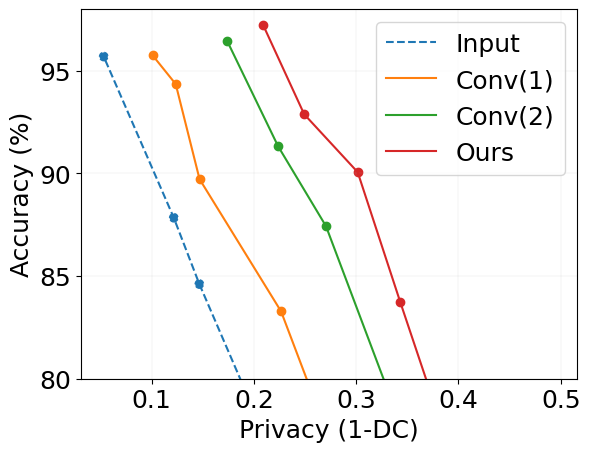}
\label{fig:ecg_acc_dc_propose1}
}
\hfill
\subfloat[Compared to others]{
\includegraphics[width=.23\textwidth]{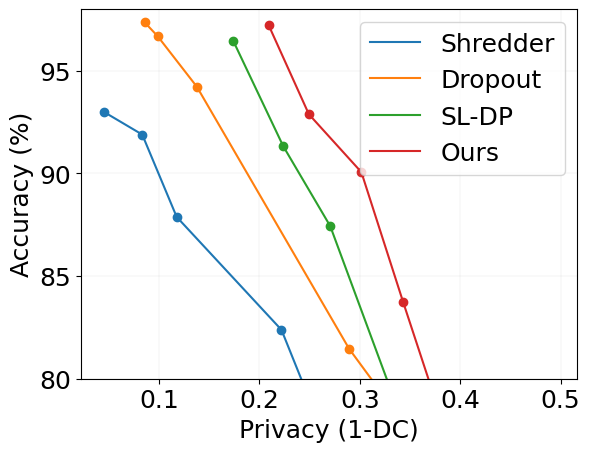}
\label{fig:ecg_acc_dc_propose2}
}
\caption{\Upd{Accuracy and privacy trade-offs of our approach on the ECG dataset.}}
\label{fig:ecg_acc_dtw_our}
\end{figure}

\Upd{In the experiments on introducing DP noise to different local layers of SL, we vary the noise levels and measure the model accuracy impacted together with the data leakage from reconstruction results. Following \cite{SLon1DCNN20Sharif}, we use the Distance Correlation (DC), a measure of dependence between two paired vectors, to quantitatively measure the data leakage. Note that DC is on the scale of $0$-to-$1$, where $0$ refers to independent vectors while $1$ means highly dependent and fully similar. Therefore, the dissimilarity, denoted by $(1-DC)$, indicates the amount of privacy preservation in our experiments. From the results plotted in Fig. \ref{fig:ecg_acc_dtw_our}, we can observe the similar trends as in our experiments with image datasets, where introducing noise to deeper layers results in higher accuracy-privacy trade-offs.}

\Upd{The additional experiments on ECG dataset validate our insights that introducing noise to the split layer result in optimal trade-offs. Note that, the trade-off can be easily controlled by the amount of noise added. Fig. \ref{fig:ecg_acc_dc_propose1} also presents the results of our proposed approach to enhance performance of SL-DP by resizing the smashed data. Following Eq. \ref{eq:upSampling}, we append a transposed 1-D convolutional layer to up-sample smashed data to match input size. We can further reduce leakage like introducing noise to Input layer, therefore, achieve the best accuracy-privacy trade-off. The enhanced privacy is achieved by paying more computation costs for processing the additional transpose layer. Note that, clients also get another benefit from reducing communication costs to the server as smaller smashed data size. However, future investigation should evaluate these costs in details to provide best practices for practitioners.}

\Upd{We also conduct further experiments to compare our approach to other obfuscating methods for privacy preserving solutions. Fig. \ref{fig:ecg_acc_dc_propose2} shows the accuracy-privacy trade-offs of Shredder \cite{Shredder20Mireshghallah}, Dropout \cite{Dropout12Hinton}, SL with DP (SL-DP), and our approach, which is built on the top of SL-DP. Shredder learns additive noise distributions to reduce the information content of smashed data without altering the model weights of clients and server. Dropout, which deactivates random neurons in one layer by setting their output to $0$, is used in \cite{AttackingProtecting22He} to provide effective defense against model inversion attacks. However, both Shredder and Dropout are designed to apply on a pre-trained network, resulting in more accuracy impacted so that their accuracy-privacy trade-offs are worse than SL-DP and our approach as demonstrated in Fig. \ref{fig:ecg_acc_dc_propose2}. Throughout our experiments with 2D-image and 1D time-series datasets, we validate the efficiency and effectiveness of our proposed approaches to enhance performance and privacy preservation of SL when introducing DP. However, exploring additional learning tasks with diverse network models, datasets, and data distributions, as well as investigating potential label leakage, opens up promising avenues for future research.}

\section{Conclusion}\label{conclusion}
This paper investigates the delicate balance between accuracy and privacy when leveraging DP to protect data privacy in SL. The study primarily explores the trade-offs between accuracy and privacy by introducing DP noise to different SL local layers with varying privacy budgets. Furthermore, based on the insights into the accuracy and privacy trade-offs, a method is proposed to reduce the size of smashed data, aiming to limit the data leakage and lower communication overhead, which is one of the notable drawbacks of SL. Expanding the scope to encompass a multi-client environment, with different privacy requirements, reveals potential reductions in accuracy among clients following training. To tackle the identified challenge, the study proposes a novel approach to enhance the learning performance by allowing the server to review clients' noise distributions during training. Through a comprehensive array of experiments utilizing real-world datasets, the results presented demonstrate a significant improvement in accuracy for all clients requiring noise, effectively aligning their performance with the benchmark reference. Examining the accuracy-privacy trade-off and introducing strategies to enhance client accuracy provide valuable insights for effectively preserving data privacy in the context of distributed learning across heterogeneous devices, such as mobiles and IoT. \Rev{Future research directions include conducting formal analysis,} exploring the optimal augmented data size to optimize server computations, 
examining other learning tasks, network models, and privacy-preserving techniques.

\bibliographystyle{IEEEtran}
\balance
\bibliography{refs}
\end{document}